\documentclass[twocolumn]{emulateapj}
\usepackage{natbib,amssymb}

\bibpunct{(}{)}{;}{a}{}{,} 
\usepackage{graphics,graphicx} 

\shorttitle{New dust components in the Red Rectangle}
\shortauthors{F.~Markwick-Kemper et al.}
\begin{document}

\title{Spitzer detections of new dust components in the outflow of the Red Rectangle}

\author{F.~Markwick-Kemper\altaffilmark{1,2}, J.D.~Green\altaffilmark{3}, E.~Peeters\altaffilmark{4}} 
\altaffiltext{1}{Department of Astronomy, P.O.~Box 3818, University of Virginia, Charlottesville, VA 22903-0818; {\tt ciska@virginia.edu}}
\altaffiltext{2}{Spitzer Fellow; Department of Physics and Astronomy, University of California Los Angeles, 475 Portola Plaza, Los Angeles, CA 90095-1547}
\altaffiltext{3}{Department of Physics and Astronomy, University of Rochester, Rochester, NY 14627}
\altaffiltext{4}{NASA Ames Research Center, MS 245-6, Moffett Field, CA 94035}

\begin{abstract}
We present Spitzer high spectral resolution IRS spectroscopy of three
positions in the carbon-rich outflow of post-AGB star HD 44179, better
known as the Red Rectangle. Surprisingly, the spectra show some strong
unknown mid-infrared resonances, in the 13--20 $\mu$m range. The shape
and position of these resonances varies with position in the nebula,
and are not correlated with the PAH features.  We conclude these
features are due to oxygen-rich minerals, located in a region which is
believed to be predominantly carbon-rich. We provide possible
explanations for the presence of oxygen-rich dust in the carbon-rich
outflows.  Simple Mg-Fe-oxides (Mg$_{1-x}$Fe$_x$O) are suggested as
carriers of these unidentified features.
\end{abstract}


\section{Introduction}

Ever after its initial discovery paper \citep{CAC_75_RR}, HD 44179,
lovingly nick-named \emph{the Red Rectangle}, has proven to be a truly
enigmatic astrophysical object, and hosts many unique phenomena. HD
44179 is a post-Asymptotic Giant Branch (AGB) star with a stellar
companion and surrounded by a circumbinary disk and a biconical
 nebula.  In the optical, HD 44179 shows the spectral
signature of the broad extended red emission
\citep[ERE;][]{SCM_80_HD44179,VCG_02_ERE}, of which the carrier is
still unknown. Superposed on this are sharper emission features which
may be the Diffuse Interstellar Bands normally seen in absorption
\citep{SWM_92_RR}. In the infrared the nebula exhibits remarkable
evidence of a mixed chemistry: spectroscopy obtained with the Infrared
Space Observatory (ISO), shows the presence of both oxygen-rich dust,
in particular silicates, and carbon-rich components in the form of
PAHs \citep{WWV_98_RedRectangle}. The authors suggest that the
oxygen-rich dust predominantly resides in the disk, while narrow band
imaging shows that the PAH emission is extended and follows the
outflows \citep{BRT_93_HD44179,SGL_93_HD44179}.  The beautiful optical
images obtained with the Hubble Space Telescope \citep{CVB_04_HST}
show that the morphology of the nebula is far from
homogeneous. Regularly spaced brightness enhancements exist, perhaps
related to the gas or dust density in the outflow. 
The ISO spectroscopy presented by
\citet{WWV_98_RedRectangle} lacks the spatial resolution to show any
correlation between the density enhancements seen in the Hubble images
and the chemical composition. In this work, we report Spitzer Space
Telescope observations of the Red Rectangle, to study spatial chemical
variations in the outflow.

\section{Observations and data reduction}
\label{sec:obs}

\subsection{Observations}

\begin{figure}
\plotone{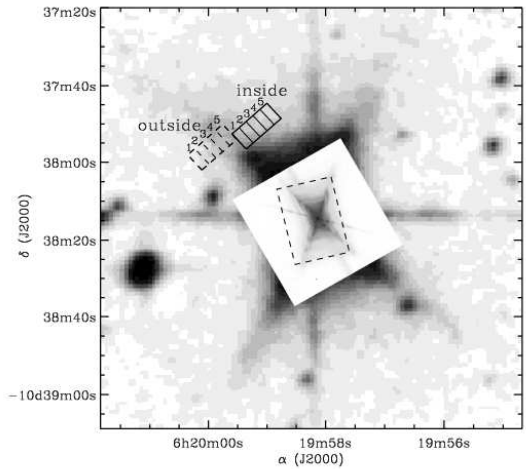}
\caption{The slit positions of the IRS SH observations on an image of
the Red Rectangle from the red band of the Digitized Sky Survey 2. 
The foreground image is taken in the WFPC2
F622W band on the Hubble Space Telescope \citep{CVB_04_HST}. The two
slit positions discussed in the text, are indicated with the labels
'inside' and 'outside' with respect to the biconical
morphology. Positions of the individual pixels in the spatial
direction within the slit positions are indicated with numbers. The dashed
box in the center indicates the aperture used with the ISO SWS observations 
\citep{WWV_98_RedRectangle}.   }
\label{fig:slitpos}
\end{figure}

We used the InfraRed Spectrograph \citep[IRS;][]{HRV_04_IRS} on
Spitzer \citep{WRL_04_Spitzer} to obtain high resolution spectroscopy
of some positions in the outflow of the Red Rectangle.  The observed
regions in the outflow are about 30$''$ away from the central
source, to avoid saturation by the central binary and its direct environment. 
Here we report the first results obtained with the Short High
(SH) module of the IRS in two lines-of-sight in the nebula; a full
description of the Spitzer data will be presented in a forthcoming
paper (Markwick-Kemper et al.~\emph{in prep.}). The first slit
position probes the region just outside the biconical X, while the
second one probes inside the biconical region
(Fig.~\ref{fig:slitpos}). The coordinates of these slit positions are
$\alpha$ = 6$h$19$m$59.94$s$, $\delta$ =
$-$10$^\mathrm{o}$37$'$56.0$''$ (J2000) and $\alpha$ =
6$h$19$m$59.18$s$, $\delta$ = $-$10$^\mathrm{o}$37$'$50.5$''$ (J2000),
respectively. At both these positions we observed 2 cycles
of 30 seconds. The SH module covers the 10.0--19.5 $\mu$m wavelength
range, with a spectral resolution of $R\approx600$.

\subsection{Data reduction and calibration}

The spectra were reduced using the {\sc smart} package
\citep{HDH_04_smart}. We performed our data reduction on the products of
 pipeline version S9.5. We worked with the \emph{droopres} images,
which is the least processed product available to the observer. The
last few processing steps of the pipeline were repeated off-line and
adjusted to the needs of our observations: First, we divided the
\emph{droopres} images by the flatfields, to get the equivalent of the
Basic Calibrated Dataproduct (BCD). We applied S10.0 flatfields,
which yielded less noisy results, to the S9.5 \emph{droopres} data;
while no S10.0 \emph{droopres} data were distributed to the
observers.  Subsequently, full aperture extraction using {\sc
smart} was performed on these images, using the flux conversion and
other calibration files provided by the Spitzer Science Center
(SSC). Bad pixels were removed by hand.   Finally, where the
orders were overlapping in wavelength coverage we cut off the band
edges according to the order optimalization provided by the SSC.  In
addition to the full aperture extraction, our data was
sufficiently complex that performing sub-aperture extraction was
necessary. This mode of extraction is not offered for the
high-resolution modules in the standard {\sc smart} version,
however,  modified the full-aperture extraction routine to extract a
specified fraction of the aperture and produce a spectrum of the flux
for that sub-aperture.  We approximated the size of the two high
resolution slits as five pixel widths in the spatial direction by two
pixel widths in the dispersion direction.  We then extracted five 
sub-apertures of an entire column in the dispersion direction
with a width of one pixels in the spatial direction.   The sum of the five 
sub-apertures is equal to the original full aperture extraction.

Currently the IRS data calibration is optimized for point sources, and
little attention has been given so far to the extraction of spectra
from extended sources. Several problems arise when extended source
calibration is attempted.  First, no extended source flux conversion
tables are available at this time. Using the flux conversion tables
for point sources affects the absolute flux levels to some extent,
although we expect the relative flux levels to be reasonably accurate.
Second, it is not possible to use the measurements of the standards
stars to calibrate the flux levels of extended sources, since it will
cause fringing. A two-dimensional spectral response function is
necessary to overcome this problem, but is not currently available.
Finally, geometric effects should be taken into account, since the
slit projection on the detectors covers only 2$\times$5 pixels, and
the point spread function is -- depending on wavelength -- about 3
pixels large, comparable in size to both the individual pixels as well
as the entire slit. \citet{SDA_04_extended} discuss the wavelength
dependence of the point spread function, causing an additional slope
in the spectra. For the high-resolution modules, the tilt of the slit
with respect to the detector combined with the small number of pixels
in the slit causes additional low-frequency fringing in the
spectra. While the first geometric effect may be easily corrected, the
second effect is fundamentally uncorrectable, since the spatial
distribution of photons on a pixel is unknown.

\begin{figure}
\plotone{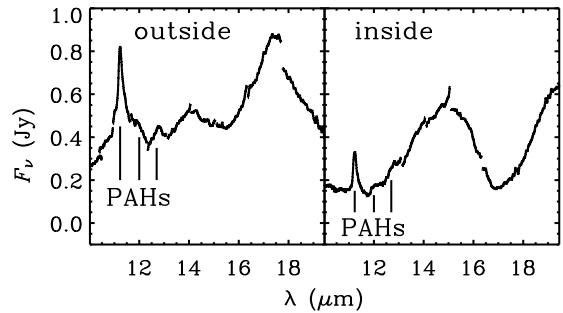}
\caption{The full aperture spectra obtained from the two slit positions,
outside and inside of the biconical region
respectively. Indicated are the presence of the out-of-plane bending
modes of PAHs \citep{HVP_01_OOPS}.  }
\label{fig:full}
\end{figure}

\begin{figure}
\plotone{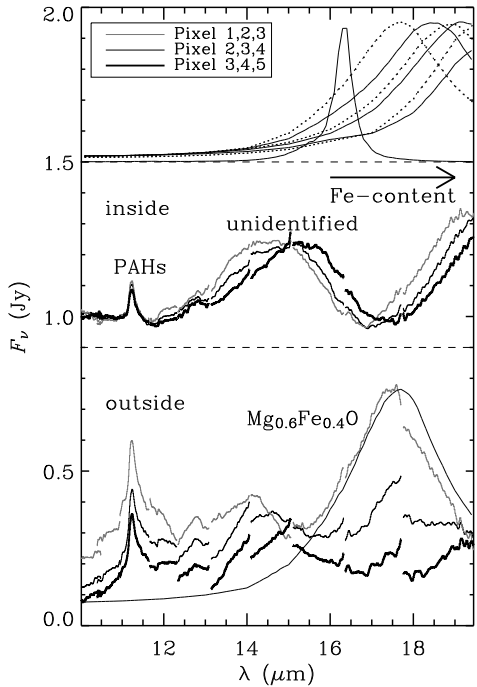}
\caption{Sub-aperture extractions compared with opacities of
Mg-Fe-oxides.  The bottom and middle parts of the plot shows 3-pixel
wide sub-aperture extractions obtained in the lines-of-sight outside
and inside the biconical region respectively. Three sub-aperture
extractions are presented, centered on pixel 2, 3, and 4. The spectra
obtained from the position inside the biconical region are offset with
0.9 Jy for clarity. The top part shows Mie opacities of Mg-Fe-oxides
\citep{HBM_95_oxide} in arbitrary units.  In increasing wavelength of
peak position and increasing Fe-content, they are the opacities of
Mg$_{1-x}$Fe$_x$O, with $x =$ 0; 0.4; 0.5; 0.7; 0.8; 0.9; and 1.
Besides the PAH features around 11 $\mu$m, two broad resonances can be
seen in the spectra from inside the biconical region; a feature from
13-16 $\mu$m and a feature peaking longward of 19 $\mu$m.  While the
feature peaking longwards of 19 $\mu$m can be fitted with a few
different Mg-Fe-oxides, the carrier of the 13--16 $\mu$m feature
remains unidentified.  The spectra in the bottom part show a strong
feature peaking at $\sim 17.5$ $\mu$m, which originates mostly from
pixel 1, as can be deduced from comparison of the sub-aperture
extractions. The opacity of Mg$_{0.6}$Fe$_{0.4}$O is overplotted. }
\label{fig:sub}
\end{figure}

Fig.~\ref{fig:full} shows the full aperture extracted spectra of the
two lines-of-sight of interest, inside and outside the biconical
region.    The observed variation in
spectral features between the two positions is very large, and
variations occur even on the scale of the pixel size. In
Fig.~\ref{fig:sub}, we show how the spectra change within both
apertures, by extracting 3 pixel wide sub-aperture spectra while
stepping along the slit one pixel at a time.

\section{Solid state features in the spectrum}
\label{sec:solid}

\subsection{PAHs?}

It is straightforward
to identify the features shortward of 13 $\mu$m present in all
lines-of-sight with the out-of-plane bending modes of the polycyclic
hydrocarbons \citep[PAHs; e.g.~][]{HVP_01_OOPS}, but these features
are dwarfed by the spectral features detected at wavelengths longer
than 13 $\mu$m. These new resonances show some interesting and unique
characteristics. First of all, the bands are very broad (several
microns) and are very strong with respect to the dust
continuum. Moreover, there is not a single set of well defined
wavelengths at which these resonances occur, but instead the peak
positions seem to vary spatially.

Interestingly, inspection of the ISO SWS spectroscopy shows no
evidence for the presence of the newly detected features, perhaps
because in ISO's large beam many lines-of-sight are averaged, and
individual contributions at various wavelengths gets averaged out to a
broad, shallow plateau, or alternatively, the carrier of these
features may only be present in detectable amounts at considerable
distance from the disk, a region not probed by ISO's line-of-sight 
(Fig.~\ref{fig:slitpos}).

Here we will explore possible carriers for the new features. Although
the outflow shows strong PAH emission at 11.2 $\mu$m even at this
large distance from the central star, we can rule out PAHs as the main
carrier.  Emission from PAHs in the 13-20 $\mu$m wavelength range is
known and is also variable \citep{HVP_01_OOPS,VHP_00_PAHplateau}. PAH
emission from 15--20 $\mu$m varies from a very broad plateau that
perhaps can be disentangled into a number of completely separate
resonances in the 15--20 $\mu$m range
\citep{VHP_00_PAHplateau,WUS_04_NGC7023,SDA_04_extended}, which do not
vary in peak position. Although these PAH bands fall in the correct
wavelength range, they are much weaker than the 11.2 $\mu$m PAH band
\citep{PMH_04_plateau}, and their FWHMs do not match.  The low
intensities and the fixed wavelengths at which the components of the
PAH plateau are found, we rule out PAHs as a significant contributor
to the resonances observed.

\subsection{Mg-Fe-oxides}

The extreme breadth of the resonances observed in the 13--20 $\mu$m
range is characteristic for
solid state carriers, and the shift in the peak positions can be
explained by variations in the temperature, composition, grain shape,
grain size or lattice structure. The strong feature-to-continuum ratio
dictates that the carrier has clean resonances without a strong
continuum component, while the breadth implies an amorphous
structure. Moreover, the opacities in the UV and optical need to be
high enough so that the grains heat to sufficiently high temperatures,
while the band strength in the material needs to be strong enough so
that the amount of dust required to explain the feature does not
violate abundance constraints.

\begin{figure}
\plotone{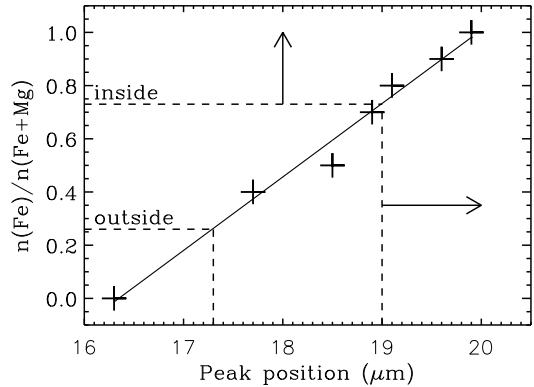}
\caption{Change of peak position with changing composition of
Mg$_{1-x}$Fe$_x$O. The pluses indicate the
the peak position as a function of composition, with an average error
bar of 0.15 $\mu$m. The number density ratio on the vertical axis is 
the Fe-content $x$ of the laboratory samples from
\citet{HBM_95_oxide}.  The solid line represents a least squares fit through
these data points. The peak position outside the biconical region is
measured and corresponds to an iron content of 26\%, whereas the
inside the biconical region, the peak positions are found longward of
19 $\mu$m (indicated with the arrow), suggesting an Fe content of 73\%
or more.}
\label{fig:fetrend}
\end{figure}

Although it is believed that the outflows of the Red Rectangle are
dominated by carbon-rich components, the features from 13--20 $\mu$m
show this picture is incomplete. Simple Mg-Fe-oxides are a possible
carrier for these new features. In 
Fig.~\ref{fig:sub} the
spectra from the region outside the biconical outflow are compared to a few
Mg-Fe-oxide minerals calculated from optical properties measured by
\citet{HBM_95_oxide}. Indeed, peak strength and width are easily
fitted, while the peak position suggests that a composition slightly
more Mg-rich than Mg$_{0.6}$Fe$_{0.4}$O is likely. The situation
inside the biconical region seems quite different. There are two clear
differences; a distinct band centered near 15 $\mu$m and a strong rise
starting near 17 $\mu$m. Interestingly, features of both bands, 
such as peak position and onset of the rise, shift as a function of
position in the slit, and the shift seems correlated. This is
consistent with a changing Fe-content in a relatively Fe-rich oxide.  
Using the correlation between composition and
peak position for spherical grains (Fig.~\ref{fig:fetrend}) it is
possible to constrain more precisely the compositions of the
oxides. We find that the oxides inside the biconical region have an
Fe-content Mg$_{<0.27}$Fe$_{>0.73}$O, while the oxides outside the
biconical region have a more Mg-rich composition of
Mg$_{0.74}$Fe$_{0.26}$O.

Taking the Mg-Fe oxides as an example, we determine that oxide grains with
a grain size of 0.1 $\mu$m heat up to about 75 K at 30$''$ of the
central star, using $R_{\ast} = 43 R_{\odot}$, $T_{\mathrm{eff}} = 7750$
K, $d = 710$ pc \citep{MST_02_RR} and assuming no extinction toward
the central star. 
To account for the observed flux levels in the features a dust mass of
$M_{\mathrm{dust}} \approx 2.4 \times 10^{28} \mathrm{g} = 1.2 \times
10^{-5} M_{\odot}$ in the form of Mg-Fe-oxides is required in the
aperture of each of the slit positions, an amount which seems large.
The distance of 710 pc derived by \citet{MST_02_RR} might be an
overestimation \citep[see e.g.~][]{JTB_97_biggrains,HTO_04_RR}, which
would affect the derived dust masses considerably. For a distance of
500 pc for instance, the dust can be heated to 150 K at 30$''$, and an
oxide mass of only $\sim 9 \times 10^{25}$ g, which is comparable to
the mass of our moon, is needed in each aperture 
to account for the observed emission features.

\subsection{The 13--17 $\mu$m band}

The strong feature at 13--17 $\mu$m remains unidentified. However,
because the peak position shifts with the change in onset of the
feature longward of 17 $\mu$m we conclude that the two carriers are
related, and that perhaps the 13--17 $\mu$m feature is caused by a
mixture of Mg-Fe-oxides with another material. A good candidate are
spinels (MgAl$_2$O$_4$), which have features in this wavelength range
\citep{FPM_01_spinels}.  Their UV/optical opacity is too low to
provide enough grain heating to explain the observed flux levels on
their own, but when they are in thermal contact with oxides, the
spinels may get warm enough to explain the observed flux level, while
the changing Fe-content still explains the shift in peak position.

\section{Discussion}
\label{sec:discussion}

The detection of new mid-infrared features in the bipolar outflow of
HD 44179 is surprising. These features are carried by oxygen-rich
minerals, possibly including simple Mg-Fe-oxides.  Previously it was
argued that the nature of the chemistry in these outflows is
carbon-rich, while the dust in the circumbinary disk is oxygen-rich
\citep{WWV_98_RedRectangle}. Independently obtained spatially resolved
ground-based observations of the PAH features in the infrared support
this view \citep[e.g.][]{KHM_99_RR,SKM_03_RR,MKO_04_RR}. Our work,
along with the recent suggestion of small PAHs in the circumbinary
disk \citep{VWG_05_smallPAHs}, indicate that the chemical distribution
to the Red Rectangle may be more complicated than previously thought;
both the disk and the outflow are not strictly oxygen-rich or
carbon-rich, and apparently mixed chemistry environments occur in both
components of the system. There are clearly different episodes in mass
ejection \citep{CVB_04_HST} and these may be connected to changes in
the chemical composition.

The origin of the oxygen-rich dust in the outflow of HD 44179 remains
speculative at best.  It is possible that the oxides found in the
outflow are remnants of an earlier, circumstellar mass loss phase,
which is currently being overtaken by the bipolar outflow.  Simple
oxides are found in the low density winds of semi-regular variables
and early type Miras \citep{PKM_02_19.5um}, likely
progenitors of a system like the Red Rectangle.  These stars are known
to develop a outflow velocity which is significantly lower than what
is seen in later stages of the mass loss phase
\citep[e.g.~][]{KM_85_massloss}.  An alternate explanation is that the
oxygen-rich dust found in the bipolar outflow originate from the circumbinary
disk. Interaction between the stellar wind and the disk may have
eroded the disk and the wind may have dragged some of the disk
material along in the collimated outflow directions. In this
interaction, the oxygen-rich dust grains from the disk
\citep[predominantly silicates;][]{WWV_98_RedRectangle} may have
partially or completely evaporated, and recondensed high above the
disk. When recondensation occurs, the density will have dropped
considerably, and the formation of significant amounts of anything
beyond the simplest condensates may have been difficult, which
could explain why the oxides are dominating the spectroscopy.

Of course, whether Mg-Fe-oxides are indeed the carriers of these newly 
discovered oxygen-rich dust features remains to be seen, and further 
studies of the spectral appearance and spatial distribution of these
features is required to further establish the nature of the dust.

\begin{acknowledgements}
FMK is grateful for her time spent at UCLA as a Spitzer Fellow,
working with Mike Jura. His detailed knowledge of the Red Rectangle
and his insightful comments were of crucial importance to this
project. We also thank Lou Allamandola for sharing his thoughts on the
origin of the spectral features discussed in this paper. We thank Dan
Watson and Ben Sargent for aiding in developing IRS data reduction
tools, Jan Cami for careful reading of the manuscript and Hans Van
Winckel for making available the HST WFPC2 image of the Red
Rectangle. Support for this work was provided by NASA through the
Spitzer Fellowship Program, under award 011 808-001; and by the National
Research Council.
\end{acknowledgements}

\end{document}